\documentstyle[prb,aps,psfig]{revtex}

\begin{document}
\twocolumn[\hsize\textwidth\columnwidth\hsize\csname %
@twocolumnfalse\endcsname
\draft
\preprint{ver. 2.2}
%
%
\title{Application of the generalized three-dimensional Jordan-Wigner 
transformation to the bilayer Heisenberg antiferromagnet}
\author{B. Bock$^1$ and M. Azzouz$^{1,2,}$\cite{email}}
\address{$^1$Department of Physics, University of Northern British Columbia, 
Prince George, British Columbia, Canada
V2N 3Z9.}
\address{$^2$Max-Planck Institut f$\ddot{u}$r Physik Komplexer 
Systeme, N$\ddot{o}$thnitzer
Str.38, 01187 Dresden, Germany}
\date{\today}
\maketitle
\begin{abstract}
We extend the definition of
the Jordan-Wigner transformation to three dimensions
using the generalization of ideas that were used in the two-dimensional 
case by one of the present authors. Under this transformation,
the three-dimensional XY Hamiltonnian is transformed into a system
of spinless fermions coupled to a gauge field with only
two nonzero components.
We calculate the flux
per plaquette for the 3 elementary perpendicular 
plaquettes of a cubic lattice,
and find that it is nonzero for only two of the plaquettes.
We provide a simple interpretation for the average
phase-per-plaquette being $\pi$ on the plaquettes where it is nonzero.
Then we apply these new findings
to the investigation of the Heisenberg bilayer antiferromagnet.
\end{abstract}
\vspace{0.2cm}
]
\narrowtext
%
%
%
%

In 1928, Jordan and Wigner\cite{jw} introduced a transformation to
write the spin operators $S^{-}$ and $S^{z}$ in
terms of Fermi operators in one dimension.  For example, this 
transformation allows a mapping 
of the XY Hamiltonian in one dimension into a 
Hamiltonian of noninteracting spinless fermions.
Trying to generalize this result to two dimensions 
proved to be difficult. But, a 
natural extension of the Jordan-Wigner (JW) 
transformation was introduced in 1993 
by one of the present authors\cite{azzouz} to study the interchain-coupling 
effect on the one-dimensional
spin-1/2 Heisenberg antiferromagnets. Several other authors 
have given implicit ways of defining a generalized 
JW transformation in two dimensions.\cite{fradkin,wang}  
The generalization of the JW transformation has been attempted 
even in three dimensions by Huerta and Zanelli\cite{zanelli} who introduced
an implicit transformation, and by Kochma\'nski\cite{koch} 
whose transformation, however, does not preserve some of the 
spin-commutation relations.

In this work, we follow the procedure of Ref.\cite{azzouz} to extend the
definition of the JW transformation in three dimensions.
This paper is organized as follows.
We start by reviewing the JW 
transformation in one and two dimensions. 
Then we define its three dimensional version. For the XY model,
we calculate the components
of the effective gauge field that results from the transformation,
and the flux
per plaquette as one spinless-fermion moves around it. Finally,
we use this transformation to study the effect of the interlayer 
coupling on the Heisenberg bilayer.

The JW transformation can be performed independent of 
any model Hamiltonian.  In one dimension this is written as
\begin{eqnarray} 
&&S_{i}^{-} = c_{i}e^{i\phi_i},\ \
S_{i}^{z} = c^{\dagger}_{i} c_{i} - 1/2.
\end{eqnarray}
The phase $\phi_i=
{\pi \sum^{i-1}_{j=0} c_{j}^\dag c_j}=
\pi \sum_{j} w(i,j)n_j$
where
\begin{eqnarray}
w(i,j)=\Theta(i-j)(1-\delta_{i,j}),
\label{w1}
\end{eqnarray}
with $\Theta(i-j)=1$ for $j\le i$ and $0$ for $j> i$
being the (discrete) Heaviside step function.
Note that the phase term in this transformation is 
obtained by summing over all 
those sites to the left of site $i$. 
Also, when we calculate
spin commutation relations we find that the fermion-number operator,
at the site with the lowest index and hosting one spin of the commutator, 
is contained only once in the
resulting phase (refer to Ref. {\cite{azzouz}} for more details).
This assures that the spin commutation relations are preserved.
In two dimensions, a generalized transformation
based on this idea has been introduced in Ref.[\cite{azzouz}].
This was written as:
\begin{eqnarray}
S_{i,j}^{-} &&= c_{i,j} \; e^{i\pi \phi_{i,j} },\ \ 
S_{i,j}^{z} = c_{i,j}^{\dagger} \; c_{i,j} - 1/2, \cr
\phi_{i,j}  &&=  \sum^{i-1}_{\alpha=0} \sum^{\infty}_{\beta=0}
n_{\alpha, \beta} + \sum^{j-1}_{\beta=0} n_{i, \beta},
\label{jw2d}
\end{eqnarray}
where now a given site is specified by two indices, $i$ 
along the x-axis and $j$ 
along the y-axis.  In this transformation, the phase at site $(i,j)$ 
is obtained by 
summing over all sites to the left of the vertical line passing
through $(i,j)$ as well as all sites 
directly below $(i,j)$. This is a
necessary and sufficient condition that allows 
the generalized JW transformation 
to preserve all spin commutation relations.
The phase term in (\ref{jw2d}) can be written as:
$e^{i\pi\sum_{{\bf z}=0}^{\infty} w({\bf x},{\bf z}) n({\bf z})}$ with
\begin{eqnarray} 
w({\bf x},{\bf z})=&&\Theta(x_{1}-z_{1}) ( 1 - \delta_{x_1,z_1}) \cr
   &&+ \Theta(x_{2} - z_{2}) \delta_{x_1,z_1} ( 1 - \delta_{x_2,z_2});
\label{w2}
\end{eqnarray}
${\bf x}=(x_1,x_2)$ and ${\bf z}=(z_1,z_2)$ denoting sites 
on the square lattice.
A necessary condition for the preservation of the spin commutation relation
is that the following condition is satisfied
\begin{eqnarray}
e^{i\pi w({\bf x},{\bf z})}=-e^{i\pi w({\bf z},{\bf x})}.
\label{ephase}
\end{eqnarray}
A relation that is indeed fulfilled by Eq.~(\ref{w2}).

Next, consider the transformation of the XY model under (\ref{jw2d}),
and calculate the flux per plaquette generated as one spinless
fermion moves around an elementary plaquette. The phase 
change is found to be:
\begin{equation} 
(n_{i,j} - n_{i+1,j})\pi. 
\label{2dplaque}
\end{equation}
This result is more general than the one of Fradkin,\cite{fradkin}
who reported that the phase should be $\pi$,
and also shows that Shaofeng's\cite{shaofeng} claim that the 
flux per plaquette is zero is not true.

Using the same procedure described above
for one and two dimensions, we define
the JW transformation in three dimensions by:
\begin{eqnarray}
S_{i,j,k}^{-} &&= c_{i,j,k}  e^{i\pi \phi_{i,j,k} },\ \ 
S_{i,j,k}^{z} = c_{i,j,k}^{\dagger} c_{i,j,k} - 1/2 \cr
\phi_{i,j,k}  &&= \sum^{k-1} _{\epsilon=0} 
	\sum^{\infty}_{\alpha=0} \sum^{\infty}_{\beta=0} n_{\alpha, 
\beta, \epsilon} 
+ \sum^{i-1}_{\alpha=0} \sum^{\infty}_{\beta=0} n_{\alpha, \beta, k}\cr
&& + \sum^{j-1}_{\beta=0} n_{i, \beta, k} 
\label{jw3d}
\end{eqnarray}
where now the phase is:
\begin{eqnarray} 
\phi({\bf x})=\sum_{z=0}^{\infty} w({\bf x},{\bf z}) n({\bf z})
\label{phase3}
\end{eqnarray}
with
\begin{eqnarray} 
w({\bf x},{\bf z}) = &&\Theta(x_{3}-z_{3}) 
( 1 - \delta_{x_3,z_3}) \cr
&& + \Theta(x_{1} - z_{1}) \delta_{x_3,z_3} ( 1 - \delta_{x_1,z_1} ) \cr
&& + \Theta(x_{2} - z_{2}) \delta_{x_1,z_1} \delta_{x_3,z_3} ( 1 - 
\delta_{x_2,z_2} ),
\label{w3}
\end{eqnarray}
${\bf x}=(x_1,x_2,x_3)$ and ${\bf z}=(z_1,z_2,z_3)$ 
are now sites on the cubic lattice.
It is straightforward to check again that Eq.\ (\ref{ephase})
is satisfied,
and that all spin commutation relations are preserved.
To obtain the phase (\ref{phase3}) we choose a plane 
going through the site where we want to define the transformation.
Here the plane chosen is the one perpendicular to the vector $(0,0,1)$.
Then we sum over those sites on this plane as in 
the two-dimensional transformation.
Finally we sum over all sites that are behind this plane.

As for the phase change that occurs when a spinless fermion completes a
motion around a plaquette, we consider the elementary
plaquettes in the $xy$, $xz$, and $yz$ planes.  
The flux per plaquette in the $xy$ plane turns out to be the same as
in two dimensions, that is
\begin{equation}
(n_{i,j,k} - n_{i+1,j,k})\pi.
\label{phaseij}
\end{equation}
For a spinless fermion starting at site $(i,j,k)$, $n_{i,j,k}=1$. Then for 
this fermion to go to site $(i+1,j,k)$, this later site has to
be empty due to Pauli exclusion principle; 
thus $n_{i+1,j,k}=0$. In this approximation one finds that the phase is
$\pi$.
For a plaquette in the $yz$ plane the result is
\begin{equation}
(n_{i,j,k+1} - n_{i,j,k})\pi.
\label{phasejk}
\end{equation}
Similarly, the phase per plaquette can be set to be approximately $\pi$.
As for the phase change around an elementary
plaquette in the $xz$ plane, it is found to be 
identically zero.

The results for these fluxes per plaquette suggest that 
as a spinless fermion moves around a given plaquette, it will couple to 
a gauge field if the plaquette belongs to plane $xy$ or $yz$,
but no coupling to a gauge field occurs in the remaining plane $xz$.
This thus suggests that the effective magnetic field to which the
spinless fermions couple has only two components.
To show this, we consider the XY model written for coupled layers:
\begin{eqnarray}
H_{XY}&&={J\over2}\sum_{i,j,k}S^-_{i,j,k} S^+_{i+1,j,k}
+ {J\over2}\sum_{i,j,k}S^-_{i,j,k} S^+_{i,j+1,k} \cr
&&+ {J_{\perp}\over2}\sum_{i,j,k}S^-_{i,j,k} S^+_{i,j,k+1} + ({\rm H.c.})
\label{bilayer}
\end{eqnarray}
where $J$ and $J_\perp$ are respectively the intralayer and interlayer
coupling constants. Under the JW transformation (\ref{jw3d}),
the Hamiltonian (\ref{bilayer}) gives
\begin{eqnarray}
H_{XY}=&&J\sum_{i,j,k}
c_{i,j,k}^{\dag}c_{i+1,j,k}e^{i\pi\nabla_i\phi_{i,j,k}} \cr
&& + J\sum_{i,j,k} c_{i,j,k}^{\dag}c_{i,j+1,k} \cr
&& + J_\perp\sum_{i,j,k} 
c_{i,j,k}^{\dag}c_{i,j,k+1}e^{i\pi\nabla_k\phi_{i,j,k}} + ({\rm H.c}).
\label{bilayer1}
\end{eqnarray}
Here $\nabla_i\phi_{i,j,k}=\phi_{i+1,j,k}-\phi_{i,j,k}$ and
$\nabla_k\phi_{i,j,k}=\phi_{i,j,k+1}-\phi_{i,j,k}$ designate
the discrete derivatives along the x and z-axes, respectively.
This effective Hamiltonian describes the motion of spinless
fermions coupled to a gauge field.
The later field is given by $A_i=\nabla_i\phi_{i,j,k}$ along the x-axis,
$A_j=\nabla_j\phi_{i,j,k}$ along the y-axis, and 
$A_k=\nabla_k\phi_{i,j,k}$ along the z-axis. This leads to
\begin{eqnarray}
&&A_i
=\pi[\sum_{\beta=0}^\infty n_{i,\beta,k} +
\sum_{\beta=0}^{j-1}(n_{i+1,\beta,k}-n_{i,\beta,k})],\cr
&&A_j=0,\cr
&&A_k
=\pi[ \sum_{\alpha=0}^{i-1} \sum_{\beta=0}^{\infty}
(n_{\alpha,\beta,k+1}-n_{\alpha,\beta,k})  \cr
&& + \sum_{\beta=0}^{j-1}(n_{i,\beta,k+1}-n_{i,\beta,k}) +
\sum_{\alpha=0}^\infty \sum_{\beta=0}^\infty n_{\alpha,\beta,k}].
\end{eqnarray}
Note that the y-component of ${\bf A}=(A_i,A_j,A_k)$ is
found to be identically zero.
The effective magnetic field is obtained through the discrete curl of
${\bf A}$, that is  
${\bf B}(i,j,k)={\bf \nabla}\times {\bf A}(i,j.k)$,
[vectors ${\bf A}$ and ${\bf B}$ depend on the position $(i,j,k)$].
The components of ${\bf B}$ which are given by
\begin{eqnarray}
&&B_i=\pi(n_{i,j,k+1}-n_{i,j,k}), \cr
&&B_j=0, \cr
&&B_k=\pi(n_{i,j,k}-n_{i+1,j,k}),
\end{eqnarray}
are consistent with the results
of the fluxes per plaquette in Eqs. (\ref{phaseij}) 
and (\ref{phasejk}), (the unit area being 1).
Therefore the effective magnetic field to which
the spinless fermions are coupled has two nonzero components only.

In the second part of this paper we apply the above findings 
to the study of the
Heisenberg bilayer antiferromagnet that has received a lot of attention
in the past few years.\cite{millis}-\cite{yu}
The use of the JW transformation and 
the bond-mean-field theory\cite{azzouz,azzouz2}
is advantageous in comparison with
the Schwinger boson approach,\cite{millis,miyazaki,ng} 
modified spin-wave approaches,\cite{chubukov}
or the bond-operator mean-field theory,\cite{matsushita,yu} 
because no constraint on the number of 
particle per site is required for the JW spinless fermions.
We would like to draw the attention to the 
bond-mean field approach\cite{azzouz,azzouz2}
which has so far given very good results in the case of 
the Heisenberg ladder,\cite{azzouz3} and coupled Heisenberg
ladders.\cite{azzouz4}
The bilayer antiferromagnet is a system that presents an
order-disorder
second order quantum transition at zero temperature, 
and may be of some relevance to
some of the high temperature
superconductors in their low-doping normal state.\cite{hida}

The Heisenberg model on a bilayer reads as follows 
\begin{eqnarray}
H&&= H_{XY} + J\sum_{i,j,k}S^z_{i,j,k}S^z_{i+1,j,k} \cr
&&+ J\sum_{i,j,k}S^z_{i,j,k}S^z_{i,j+1,k}
+ J_{\perp}\sum_{i,j,k}S^z_{i,j,k}S^z_{i,j,k+1}.
\label{bilayer2}
\end{eqnarray}
Hamiltonian (\ref{bilayer2}) is simplified using the approximation where
the average flux per elementary plaquette is $\pi$ on the $xy$
and $yz$ planes, and zero on the $xz$ plane (as discussed above).
To achieve this, we choose the following configuration:
the phases are alternated $...\pi-0-\pi-0...$ along the adjacent bonds
on the y-axis, and zero on all bonds on the remaining axes.
Note that the alternated phases could be put rather on the x-axis
without changing the physical results because of gauge invariance.
In the case of the bilayer system, the alternated phases cannot however be put
on the z-axis because there is only one bond along that axis.
Further simplification is done by using the bipartite character
due to antiferromagnetic correlations.
Then we decouple the interacting quartic 
terms by introducing the alternated magnetization
parameter $m_i=2\langle S_{i,j,k}^z\rangle
=2\langle c_{i,j,k}^{\dag}c_{i,j,k}\rangle - 1=(-1)^im$,
and the bond parameters $Q=\langle c_{i,j,k}c_{i+1,j,k}^{\dag}\rangle=
\langle c_{i,j,k}c_{i,j+1,k}^{\dag}\rangle$ within the planes
and $P=\langle c_{i,j,k}c_{i,j,k+1}^{\dag}\rangle$ 
perpendicular to the planes.
The Hamiltonian takes the following form in the reciprocal space:
\begin{eqnarray}
H&&=\sum_{\bf k}(2J+J_\perp)m(e_{\bf k}^{\dag}e_{\bf k}
-f_{\bf k}^{\dag}f_{\bf k}) \cr
&& + \sum_{\bf k}[
iJ_1\sin k_x + \gamma(k_y,k_z)]e_{\bf k}^{\dag}f_{\bf k} + {\rm H.c.}
\label{bilayer3}
\end{eqnarray}
where $\gamma(k_y,k_z)=J_1\cos k_y + 
J_{\perp1}\cos k_z$, $e_{\bf k}=c_{\bf k}^A$ (resp. $f_{\bf k}=c_{\bf k}^B$) 
is the fermion operator on the sublattice A (resp. B). The quantities $J_1$
and $J_{\perp1}$ are given by $J_1=J(1+2Q)$
and $J_{\perp1}=J_\perp(1+2P)$.
The energy spectrum takes the expression:
\begin{eqnarray}
E_\pm(k)=\pm\{(2J+&&J_\perp)^2m^2 \cr
&& +J_1^2\sin^2k_x
+\gamma^2(k_y,k_z)\}^{1/2}
\end{eqnarray}
leading to the ground-state energy per site:
\begin{eqnarray}
E_{GS}={{2J+J_\perp}\over4}m^2  + 2JQ^2 + J_{\perp} P^2
 -\int{(d^3k)\over2}E_+
\end{eqnarray}
where by definition $(d^3k)\equiv{d{\bf k}/{(2\pi)}^3}$.
Because of the periodic boundary conditions 
along the z-axis, $J_\perp$ is counted twice. Thus
we have to divide the value of $J_\perp$
by 2 in all our equations, or multiply $J_\perp$ by two in the final 
results wherever $J_\perp$ appears.
\begin{figure}
\centerline{\psfig{figure=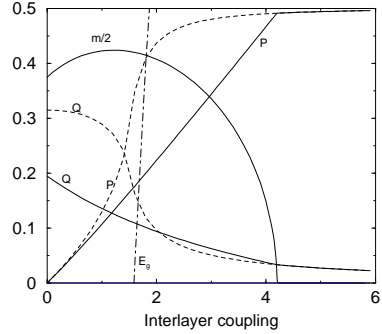,height=4.5cm,angle=0}}
\caption{The parameters $m$, $Q$ and $P$ are plotted
as a function of $J_\perp$. Full lines are 
obtained for finite $m$, and dashed lines are for $m=0$.
The dashed-dotted line is $E_g$ defined by
Eq. (\ref{gap}).}
\label{MPQ}
\end{figure}
The parameters $m$, $Q$, and $P$ are calculated self-consistently
through the equations:
\begin{eqnarray}
&&m=\int{(d^3k)}{m(2J+J_\perp)/E_+} \cr
&&Q=\int{(d^3k)}{
\{J_1\sin^2k_x  + \gamma(k_y,k_z)\cos k_y\}/4E_+} \cr
&&P=\int{(d^3k)} \gamma(k_y,k_z)\cos k_z/2E_+,
\label{mfe}
\end{eqnarray}
obtained by minimizing the ground-state energy with respect to $m$, $Q$
and $P$.
Two solutions for the 
magnetization $m/2$
are possible depending on 
the value of $J_\perp$, Fig. \ref{MPQ}.
Let us first analyze the trivial solution $m=0$ of Eqs.~(\ref{mfe}). 
In this case the energy spectrum
$E_k=\pm\{J_1^2\sin^2k_x +\gamma^2(k_y,k_z)\}^{1/2}$
presents an energy gap at ${\bf k}=(k_0,\pi,0)$ or 
$(k_0,0,\pi)$, ($k_0=0,\ \pi$):
\begin{eqnarray}
E_g=|J_{\perp1}-J_1|=|J_\perp(1+2P)-J(1+2Q)|,
\label{gap}
\end{eqnarray}
if $J_{\perp1}>J_1$. This condition is deduced from
$\gamma(k_y,k_z)=0$
which is essential for a vanishing gap. Numerically we find that 
the value above which
an energy gap opens is $J_\perp=2\times0.78J=1.56J$, see Fig. \ref{MPQ}.
However, this does not mean that this is the critical value for 
the order-disorder second-order quantum transition. 
As for the finite $m$ solution,
figure \ref{MPQ} shows that $m\ne0$ 
for $J_\perp<J_{\perp c}=2\times2.1J=4.2J$. 
The phase is ordered  antiferromagnetically
for $J_\perp<J_{\perp c}$, whereas it is a disordered quantum state for 
$J_\perp>J_{\perp c}$.
The value $J_{\perp c}=4.2J$ compares well to the result ($4.48J$)
of the Schwinger
boson approach\cite{millis,miyazaki,ng} and that ($4.3$) of the 
self-consistent spin-wave theory.\cite{hida2}
But it disagrees with the Monte Carlo simulations \cite{sandvik2} 
result ($2.5J$)
or the series expansion\cite{hida} calculation ($2.56J$).
Chubokov and Morr\cite{chubukov} reported that
while transverse spin-wave excitations are gapless (Goldstone modes)
longitudinal spin-wave excitations are gapped for $J_{\perp c}>J_\perp>0$.
Our result seems to indicate that a finite value of the interlayer coupling
($J_\perp=1.56J$) would be needed for longitudinal fluctuations to become
relevant. Note however that we have to consider gaussian 
fluctuations about our mean-field point
to account for the spin-wave excitations in the ordered phase.
We expect that fluctuations beyond the mean-field point will lead to
a better estimate of $J_{\perp c}$.
The magnetization increases with $J_\perp$, passes through a maximum,
then decreases in agreement with previous results.\cite{chubukov}

As for the parameters $P$ and $Q$, when $J_\perp$ increases
$P$ saturates at  $0.5$,
whereas $Q$ decreases to 0, Fig.~\ref{MPQ}. For very large values
of $J_\perp$ this approach becomes less accurate.
Perturbation expansion about the transverse dimers may be used.\cite{weihong,hida}
\begin{figure}
\centerline{\psfig{figure=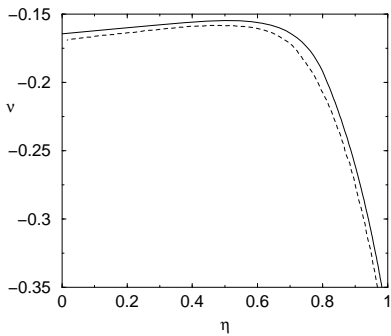,height=4.5cm,angle=0}}
\caption{The reduced ground-state energy is plotted as a function of
the reduced interlayer coupling
as defined in the text. The dashed line is from Ref. [\cite{weihong}].}
\label{GSEN}
\end{figure} 
Finally, the result of the calculation of the ground-state energy, $E_{GS}$, 
is reported in Fig.~\ref{GSEN} in order to support the validity of 
the present approach. To 
compare with the result of Weihong\cite{weihong} obtained using
Ising and dimer expansions, $\nu=E_{GS}/(4J+J_\perp)$ 
is drawn as a function of $\eta=J_\perp/(J+J_\perp)$. The maximum of
$\nu$ at ${\eta}=0.5$ coincides with that reported by Weihong,
and in general both results agree very well.

In summary, we  extended the definition of the JW
transformation to three dimensions and gave a full analysis
of its consequences in the case of the XY model. This model
is found to be transformed
to a system of spinless fermions that are coupled to a gauge field
with two components only (one of the
three components being zero).
Accordingly, the flux per plaquette
is nonzero for 2 elementary perpendicular
plaquettes, but is zero for the third one. We gave a simple interpretation
of the average flux being $\pi$ for those plaquettes where it is nonzero.
These findings were used to study the Heisenberg bilayer antiferromagnet
within the bond-mean field theory.
We recalculated the order-disorder critical coupling and the 
ground-state energy.
Very good agreement with previous results
is found. Finally,
the present three-dimensional JW transformation can be applied to 
models of three-dimensional quantum spin systems.
For simplicity a cubic Bravais lattice was used in this work,
but the transformation introduced here can be easily generalized to other
Bravais lattices.

B.B. aknowledges receipt of an undergraduate NSERC award.
M.A. is grateful to Prof. Shegelski for his support and the useful 
discussions they had together, and to Prof. Hussein for his
support.
He thanks Prof. Thalmeier and Dr. Yuan for the helpful discussions
they had during his stay at the Max-Planck Institute in Dresden.
He is also grateful to this institute for the financial support.


\begin{references}
%
\bibitem[*]{email} Correspondance could be sent to azzouzm@unbc.ca
%
\bibitem{jw} P. Jordan and E. Wigner, Z. Physik {\bf 47}, 631 (1928)
%
\bibitem{azzouz} M. Azzouz, Phys. Rev. B {48}, 6136 (1993)
%
\bibitem{fradkin} E. Fradkin, Phys. Rev. Lett. {\bf 63}, 322 (1989)
%
\bibitem{wang} Y. R. Wang, Phys. Rev. B {bf 46}, 151 (1992),
Phys. Rev. B {\bf 45}, 12604 (1992), and Phys. Rev. B 
{\bf 43}, 3786 (1991)
%
%
\bibitem{zanelli} L. Huerta and J. Zanelli, Phys. Rev. Lett. {\bf 71}, 362 (1993)
%
\bibitem{koch} M. S. Kochma\'nski, J. Tech. Phys. {\bf 36}, 485 (1995)
%
%
\bibitem{shaofeng} W. Shaofeng, Phys. Rev. E {\bf 51}, 1004 (1995)
%
%
%
%
\bibitem{millis} A. J. Millis and H. Monien, Phys. Rev. Lett. {\bf 70},
2810 (1993) and
 Phys. Rev. B {\bf 50}, 16606 (1994)
%
\bibitem{sandvik} A. W. Sandvik and D. J. Scalapino, Phys. Rev. Lett.
{\bf 72}, 2777 (1994)
%
\bibitem{sandvik2} A. W. Sandvik, A. V. Chubukov, and S. Sachdev, Phys. Rev. B
{\bf 51}, 16483 (1995)
%
\bibitem{chubukov} A. V. Chubukov and D. K. Morr, Phys. Rev. B {\bf 52}, 3521
(1995)
%
\bibitem{sandvik3} A. W. Sandvik and D. J. Scalapino Phys. Rev. B {\bf R526}
1996
%
\bibitem{miyazaki} T. Miyazaki, I. Nakamura, and D. Yoshioka, Phys. 
Rev. B {\bf 53}, 12206 (1996)
%
\bibitem{ng} K.-K. Ng, F. C. Zhang, and M. Ma, Phys. Rev. B {\bf 53}, 12196
(1996)
%
\bibitem{hida2} K. Hida J. Phys. Soc. Jpn. {\bf 59}, 2230 (1990). Note that
a modified spin wave theory (Ref.\cite{chubukov}) gives $J_{\perp c}=2.73J$.
%
\bibitem{weihong} Z. Weihong, Phys. Rev. B {\bf 55}, 12267 (1997)
%
\bibitem{matsushita} Y. Matsushita, M. P. Gelfand, and C. Ishii,
J. Phys. Soc. Jpn. {\bf 68},247 (1999)
%
\bibitem{yu} D.-K. Yu, Q. Gu, H.-T. Wang, and J.-L. Shen, Phys. Rev. B 
{\bf 59}, 111 (1999)
%
\bibitem{azzouz2} M. Azzouz and C. Bourbonnais, Phys. Rev. 
B {\bf 53}, 5090 (1996)
%
\bibitem{azzouz3} M. Azzouz, L. Chen, and S. Moukouri, Phys. Rev. 
B {\bf 50}, 6233 (1994)
%
\bibitem{azzouz4} M. Azzouz, B. Dumoulin, A. Benyoussef, Phys. Rev.
B {\bf 55}, R11957 (1997)
%
\bibitem{hida} K. Hida, J. Phys. Soc. Jpn. {\bf 61}, 1013 (1992)
%

\end{references}
\end{document}